\documentclass[12pt,a4paper,oneside]{article}
\usepackage{ifpdf}
\usepackage{a4wide}
\usepackage{amsmath}
\usepackage{graphicx}
\usepackage{rotating}
\usepackage[numbers,sort&compress]{natbib} 
\usepackage{amssymb}
\begin{document}

\textheight 21.0cm
\textwidth 16cm
\sloppy
\oddsidemargin 0.0cm \evensidemargin 0.0cm
\topmargin 0.0cm

\setlength{\parskip}{0.45cm}
\setlength{\baselineskip}{0.75cm}



\begin{titlepage}
\setlength{\parskip}{0.25cm}
\setlength{\baselineskip}{0.25cm}
\begin{flushright}
DO-TH 09/14\\
\vspace{0.2cm}
September 2009
\end{flushright}
\vspace{1.0cm}
\begin{center}
\Large
{\bf Parton Models and the Elastic Scattering of Supersymmetric
Dark Matter off Nucleons}
\vspace{1.5cm}

\large
M.\ Gl\"uck and E.\ Reya
\vspace{1.0cm}

\normalsize
{\it Universit\"{a}t Dortmund, Institut f\"{u}r Physik}\\
{\it D-44221 Dortmund, Germany} \\

\vspace{1.5cm}
\end{center}

\begin{abstract}
\noindent 
The consequences of a flavor broken SU$(3)_f$ dynamical parton model
are studied in the context of the elastic scattering of supersymmetric
dark matter off nucleons.  It is shown that the predictions of this 
model differ considerably from those of the commonly applied `standard'
parton model. Some notable properties of the dynamical parton model
are finally presented and discussed. 
\end{abstract}
\end{titlepage}


The dominant spin-dependent (SD) cross section for the elastic 
scattering of supersymmetric dark matter (DM) off nucleons is given
at the leading order (LO) of perturbative QCD and the zero DM 
velocity and momentum-transfer limit by
\begin{equation}
\sigma_{\rm SD}^N =\sigma_0\left(\sum_{q=u,d,s}\, a_q\, \Delta q^N\right)^2
\end{equation}
with $\sigma_0$ and $a_q$ specified, for example, in \cite{ref1}
and where
\begin{equation}
\Delta q^N 
  = \int_0^1 dx \left[ \delta q^N(x,Q^2)+\delta\bar{q}^N(x,Q^2)\right]\, .
\end{equation}
The ${\delta\stackrel{(-)}{q}}\,^N(x,Q^2)$ are the polarized parton 
distributions of the scattering nucleon and at LO the $\Delta q^N$
are $Q^2$-independent. 
We shall consider the ratio 
$R_{\rm SD}=\sigma_{\rm SD}^n/\sigma_{\rm_{SD}}^p$
whose dependence on  higher order QCD and form factor corrections is
suppressed.  
Its determination is furthermore independent of the poorly determined
local DM density which yields the actually registered event rates
\cite{ref1}.
For illustrative purposes we shall consider a model where 
$a_d = a_s = -a_u$, corresponding to $Z^0$-exchange dominated 
Higgsino-nucleon scattering.  
Here the isospin relations $\Delta d^n= \Delta u^p\equiv \Delta u$
and $\Delta u^n =\Delta d^p\equiv \Delta d$ yield
\begin{equation}
R_{\rm SD} 
=\left( \frac{\Delta u-\Delta d+\Delta s}
{\Delta u-\Delta d-\Delta s}\right)^2\, ,
\end{equation}
showing explicitly the sensitivity to $\Delta s$ as noticed in
\cite{ref1}.

To study this issue in more detail we compare two plausible models for
$\Delta q$.
The first is the commonly considered `standard' parton model approach,
implemented for example in \cite{ref2}, where
\begin{eqnarray}
A_3 & \equiv & \Delta u -\Delta d = F+D = 1.269\pm 0.003\nonumber
\\
A_8 & \equiv & \Delta u +\Delta d -2\Delta s = 3F-D = 0.586\pm 0.031\, ,
\end{eqnarray}
while the second `dynamical valence' parton model (DVM) \cite{ref3},
inspired by Lipkin \cite{ref4}, is based on $(q_v\equiv q-\bar{q})$
\begin{eqnarray}
\Delta u -\Delta d & = & F+D\, ,\nonumber
\\
\Delta u_v +\Delta d_v & = & 3F-D\, ,
\end{eqnarray}
and a broken SU$(3)_f$ relation $\Delta s(Q_0^2)=0$.
Some more specific features of this latter model are
\begin{equation}
\delta s(x,Q_0^2) =\delta \bar{s}(x,Q_0^2) = 0
\end{equation}
and a  `Pauli--blocking' relation
\begin{equation}
\delta u(x,Q_0^2)\,  \delta\bar{u}(x,Q_0^2) = \delta d(x,Q_0^2)\, 
  \delta\bar{d}(x,Q_0^2)
\end{equation}
where $Q_0^2= 0.26$ GeV$^2$ is the low input scale for the dynamical
parton model \cite{ref3} at LO.  In this DVM $\Delta s(Q^2)=0$ at LO
implying $R_{\rm SD}^{\rm LO}=1$ for {\em any} $\Delta u$ and 
$\Delta d$ with a similar result at the NLO where \cite{ref3}
$\Delta s(Q^2) = {\cal{O}}(10^{-3})$. 
Furthermore, it is interesting to note that the hadron level result,
$R_{\rm SD}^{\rm hadr.}=(I_3^n/I_3^p)^2=1$, coincides with the LO DVM
prediction.

In the standard parton model approach where \cite{ref2} $\Delta s\simeq -0.1$
one obtains, due to (4), a quite {\em different} result:
\begin{equation}
R_{\rm SD} = \left( \frac{1+(\Delta s/A_3)}{1-(\Delta s/A_3)}\right)^2
  = 1+4(\Delta s/A_3)+ {\cal{O}}[(\Delta s)^2] \simeq 0.7
\end{equation}
which also explicitly demonstrates the enhanced sensitivity to $\Delta s$
already noticed \cite{ref1} in the context of more realistic supersymmetric
models whose $a_q$ parameters depend on the specific input parameters
of these models.

As noticed, the hadronic uncertainties in the $R_{\rm SD}$ calculations
are dominated by the error estimates for $\Delta s$. Due to the 
limited $x$--range of the actual measurements these errors can be only
specified for the truncated moment and the authors of \cite{ref2}
quote
\begin{displaymath}
\int_{0.001}^1 dx\, \delta s(x,Q^2=10\, {\rm GeV}^2) = 
   -0.006 \pm 0.012
\end{displaymath}
and an {\em extrapolated} full moment 
$\int_0^1 dx\, \delta s(x,Q^2 = 10\, {\rm GeV}^2) = -0.05$
whose error can obviously not be estimated reliably.
This precarious extrapolation should be contrasted with the 
extrapolations for the dominant $\delta(u+\bar{u})$ and
$\delta(d+\bar{d})$ where the difference between the truncated and
full moments is much less dramatic \cite{ref2}.  The hadronic
uncertainties of the calculated $R_{\rm SD}$ in the DVM are thus
under control and expected to be small in contrast to the situation
in the `standard' parton model approach.

It is interesting to note that the DVM {\em predicted} \cite{ref3}
a behavior of $\delta s(x,Q^2)$, $\delta{\bar{u}}(x,Q^2)$ and
$\delta\bar{d}(x,Q^2)$ at $Q^2>2$ GeV$^2$ that was later confirmed
by the detailed analysis in \cite{ref2}.  In its unpolarized version
\cite{ref5} it also predicted \cite{ref6} a small--$x$ behavior of
$F_2(x,Q^2)$ and $g(x,Q^2)$ which was subsequently experimentally
confirmed at HERA \cite{ref7,ref8}.  
The broken SU$(3)_f$ relation $\stackrel{(-)}{s}\!\!(x,Q_0^2)=0$ of the
unpolarized version of the dynamical parton model was also found 
\cite{ref9} to be consistent with determinations of 
$\stackrel{(-)}{s}\!\!\!(x,Q^2 > Q_0^2)$ via neutrino-nucleon scattering
experiments \cite{ref10}.  
Having adopted $\stackrel{(-)}{s}\!\!(x,Q_0^2)=0$ also later on in
\cite{ref11} we were compelled to choose $\delta\!\!\stackrel{(-)}{s}
\!\!(x,Q_0^2)=0$ in \cite{ref3} due to the positivity constraints 
$|\delta\!\!\stackrel{(-)}{q}\!\!(x,Q^2)|\leq \stackrel{(-)}{q}\!\!(x,Q^2)$
and the fact that we used the unpolarized distributions of 
\cite{ref11} for evaluating the spin-asymmetries in \cite{ref3}.  
It has recently also been shown \cite{ref12} that the DVM model is
not only compatible with all modern data on inclusive deep inelastic
scattering and dilepton and high-$E_T$ jet production, but more
importantly that its uncertainties in the small-$x$ region are
smaller than their standard parton model counterparts due to the 
valencelike distributions of the DVM at the input scale $Q_0^2$ which
strongly constrain their small-$x$ behavior at $Q^2>Q_0^2$.

To summarize, we have shown that the cross sections and hadronic
uncertainties for the dominant spin-dependent scattering of 
supersymmetric dark-matter off nucleons are strongly dependent on
the underlying model for the polarized parton distributions. In 
particular we have pointed out that the hadronic uncertainties within
the framework of a dynamical valence model are expected to be smaller
than those in the commonly considered \cite{ref1,ref2} standard
parton model approach.  Finally some notable properties of the
dynamical valence model were presented and discussed.
\vspace{0.35cm}

This work has been supported in part by the  `Bundesministerium 
f\"ur Bildung und Forschung', Berlin/Bonn.

\newpage
\end{document}